# Generative AI: A Pix2pix-GAN-Based Machine Learning Approach for Robust and Efficient Lung Segmentation

Sharmin Akter
*Biomedical Engineering*
*Jashore University of Science and Technology*
Jashore, Bangladesh
sharmintalukder120@gmail.com

Name
*dept. name*
*(of Affiliation)*
University
*(of Affiliation)*
Newyork, USA
XXX@gmail.com

Name
*dept. name*
*(of Affiliation)*
University
*(of Affiliation)*
Newyork, USA
XXX@gmail.com

*Abstract*— Chest radiography is climacteric in identifying different pulmonary diseases, yet radiologist workload and inefficiency can lead to misdiagnoses. Automatic, accurate, and efficient segmentation of lung from X-ray images of chest is paramount for early disease detection. This study develops a deep learning framework using a Pix2pix Generative Adversarial Network (GAN) to segment pulmonary abnormalities from CXR images. This framework's image preprocessing and augmentation techniques were properly incorporated with a U-Net-inspired generator-discriminator architecture. Initially, it loaded the CXR images and manual masks from the Montgomery and Shenzhen datasets, after which preprocessing and resizing were performed. A U-Net generator is applied to the processed CXR images that yield segmented masks; then, a Discriminator Network differentiates between the generated and real masks. Montgomery dataset served as the model's training set in the study, and the Shenzhen dataset was used to test its robustness, which was used here for the first time. An adversarial loss and an L1 distance were used to optimize the model in training. All metrics, which assess precision, recall, F1 score, and Dice coefficient, prove the effectiveness of this framework in pulmonary abnormality segmentation. It, therefore, sets the basis for future studies to be performed shortly using diverse datasets that could further confirm its clinical applicability in medical imaging.

*Keywords—GAN, Generative AI, Lug Segmentation, X-ray, Shenzhen, Unet.*

## I. Introduction

Environmental factors, along with air pollution, radon gas, and other chemicals resulting from industries, have raised the incidence of lung infections and placed them as the foremost reason for mortality and morbidity in the world [1]. There is a dire need to quickly and effectively illnesses including COVID-1, pneumonia, asthma, and TB. CXRs and counted tomography image modalities help the deadliness of important information concerning lung conditions and sickness severity [2].

Of the medical imaging strategies, X-ray remains among the most commonly used due to its wide availability, low cost, non-invasiveness, and easy image acquisition [3-4]. Chest radiography forms a pivot in the diagnosis of pulmonary diseases. The bulk of the daily CXR images makes manual analysis challenging, especially considering the shortage of expert radiologists. Automation of lung segmentation from CXR images is thus primordial for efficient and effective early disease diagnosis. While several ML and image processing methods have been invented for the same purpose, most lack accuracy.

It is difficult to segment the lungs from CXR images because to the coexistence of several organs, margins, rib cages, and differences in lung size and form depending on age and gender. In addition, variation in the characteristics of various TB lesions adds to the complexity of its rapid and accurate examination [5]. Computer-aided testing can bring considerable increments in physicians' accuracy of chest X-ray screening. Integrating digital image processing and computer vision with medicine was one such revolutionary step that helped automate lung region segmentation, hence saving labour costs. The underlying principle of segmentation is learned from classifying and grouping pixel points in the same class. Deep learning methodologies have radically improved the accuracy of semantic segmentation algorithms [6]. It proposes a method for enhancing the robustness of lung segmentation with a robust pix2pix generative adversarial network.

In this work, the proposed method will be trained on the Montgomery County dataset and tested on a 20% subset of the dataset. Further evaluation regarding the model's robustness will be done on the Shenzhen dataset. The approach uses GAN techniques to increase the dependability and accuracy of lung segmentation from CXR images. This will help surmount the limitations prevailing in previous methods.

The following are the research's main contributions-

- Proposing pix2pix-GAN model for lung Segmentation.
- Augmentation of dataset.
- Robustness investigation by another dataset.
- Extensive evaluation and analysis of the model.

A review of the literature is included in Section II. Section III provides a description of the materials and approach. The outcome and discussion are presented in Section IV. Section V provides illustrations for the conclusion and next steps.

## II. Litterature Review

Machine learning is an emerging methodology for diagnosing diseases. Numerous studies have been conducted in medical diagnosis and identification [7-8]. Similarly, deep learning has been improving the diagnosis and detection of medical disorders from radiological images. Lung segmentation for diagnosing and facilitating the detection of various diseases is also a growing field.

A novel technique for the lung segmentation of thoracic CT images was developed by Caixia Liu et al. [9] utilizing a multi-scale grey correlation-based approach and a dilated U-Net model. The algorithm aims to educate lung regions, capture preliminary contours of nodules, and further refine them. The experimental findings yielded the highest possible Dice similarity coefficient above current methods, with an average of 72.14%. Radiologists can use the technique to help assess lung nodules and give appropriate treatment solutions. The PulmonU-Net protocol, which was introduced by Mary Shyni et al. [10], made use of PulmonNet modules to highlight the locations of infection on chest X-ray images. Their model created dense feature maps with leaky ReLU activation, enabling neurons to function continuously, while taking use of the global and local properties of the chest X-ray pictures. The dice similarity coefficient rose to 94.25% when the vanishing gradients were resolved. Experiments on prediction visualization and real-time testing amply demonstrated its efficacy in autonomously classifying lung infections.

To divide the lungs from chest radiographs, Saurabh et al. [11] employed a modified Unet-based deep learning architecture at many scales. The technique was evaluated for several quantitative characteristics, including accuracy, recall, precision, F1-score, and Jaccard index. It is also more power-efficient than previous deep learning techniques. The measurement's accuracy was 97.34%.

Peilin Li et. al. [12] proposed an enhanced DeeplabV3+ model for chest X-ray semantic segmentation. In this context, lung segmentation was precisely marked by various health issues that challenge accurate lung segmentation. To get around the grid effect, they combined a Vision-Transformer network with an improved backbone network that employed unlabeled chest X-ray images and a hybrid expansion convolution approach. To increase the lung edge segmentation's accuracy, they employed an edge detection operator. For the image segmentation dataset containing chest X-ray pictures, our method received a score of 93.47% mIoU on Kaggle.

Using chest X-ray images, Adel Sulaiman et al. [13] suggested a convolutional neural network architecture for lung segmentation. This model extracts relevant information from the picture by using a transpose layer and concatenate blocks. It was evaluated on five distinct data subsets, and it was trained using k-fold validation. Performance metrics, including dice coefficient, accuracy, and IoU, were calculated and yielded values of 0.96, 0.93, and 0.97, respectively.

Wufeng Liu et al. [14] refined the automated model of lung segmentation in frontal chest radiograph images of CXR by employing a pre-trained U-Net network using the LeakyReLU activation function as the decoder and Residual block with Efficientnet-b4 as the encoder. With less gradient instability, it can effectively extract lung field characteristics and enhance CXR lung segmentation with a smaller standard deviation and better resilience. The model reached an accuracy rate of 98.9% against the MC dataset. Another limitation of this study is that the investigation of the response of the created model to another dataset needed to be checked.

Ankit Misra et al. [15] presented the lung segmentation model for soft tissue and boundary identification from infected chest radiographs. With a testing dataset with spatial and reverse attention layers, it archived an accuracy of 97.12%. It outperformed architectures such as Pyramid Scene Parsing Network, U-Net, U-Net++, and Fully Convolutional Network. Wimukthi et al. [16] presented a CXR-segmented dataset containing 243,324 frontal views of CXR images and segmented masks, derived from the MIMIC-CXR dataset. Images in it were segmented using a Spatial attention architecture in U-Net, SA-UNet, and dice similarity coefficients of 96.80% and IoU of 91.97% were obtained for the segmentation of lungs. This shows that CXLSeg CXR radiographs, which are segmented, already optimize the visual feature extraction process, thereby outperforming the original MIMIC-CXR images.

Literature review shows that researchers have made much effort to develop different techniques that segment the lungs from radiography images; however, these methods need more performance and robustness issues. Most of the studies employed deep learning techniques, including U-Net and U-Net++. Notably, GAN techniques have not yet to be used in lung segmentation. This robustness was never demonstrated with a hybrid dataset where one dataset would be used for training and another for testing. In this work, we have applied a pix2pix generative adversarial network to improve lung segmentation with more robustness. We have trained on the MC data set and tested it with a 20% test set from the same dataset. Moreover, the robustness has been tested on the Shenzhen dataset.

III. MATERIALS AND METHODOLOGY

This section magnifies the dataset description and data preprocessing techniques and details the proposed model for efficient and robust lung segmentation. It integrates several strategies for accurately and effectively segmenting pulmonary abnormalities from chest X-ray images.

*A. Dataset Description*

Datasets in this research, originating from Shenzhen and Montgomery, integrate some handy resources available for studying tuberculosis using Chest X-ray imaging [17]. In cooperation with the Shenzhen Chest X-ray Database was created by the Shenzhen No. 3 People's Hospital and the National Library of Medicine, which contains 336 cases of manifestations of tuberculosis and 326 standard cases. The images were all acquired on Philips DR Digital Diagnose systems in an outpatient setting. In the Montgomery County X-ray Set, there are 138 posterior-anterior X-rays including 80 standard cases and 58 aberrant instances that show signs of TB. Every picture is in DICOM format and has been anonymized., but the radiology readings are maintained in the text files. Fig. 1 shows the images of Dataset.

*B. Data Augmentation*

Augmentation is the strategy of diversifying a dataset that impacts a betterment in the proficiency of generalization and improvement in the solidity of the machine learning model through avoidance of overfitting. The following are the augmentation techniques that must be used in this research: feature-wise centre = False, samplewise centre = False. These ensure that features and samples are not centered; hence, the nature of data remains in its original distribution. rotation_range=5 provides slight rotations to familiarize the model with the recognition of an object from a different angle. Range for width shift=0.05 and range of height shift =0.05 provide horizontal and vertical shifts from which the model fits much better the positional variations of a targeted object. shear_range=0.01 represents shearing transformation, making this model robust to distortion. zoom_range= [0.8,1.2] applies to zooming and makes the model concentrate on various parts of an image. horizontal_flip=True and vertical_flip=False flip images horizontally, increasing the dataset's diversification and helping the model learn invariant features. fill_mode='nearest' ensures that new pixels are added during transformations because such methods are filled with the closest original pixel. All these techniques increase the variability of the dataset to quite an extent, hence improving generalization ability across different scenarios.

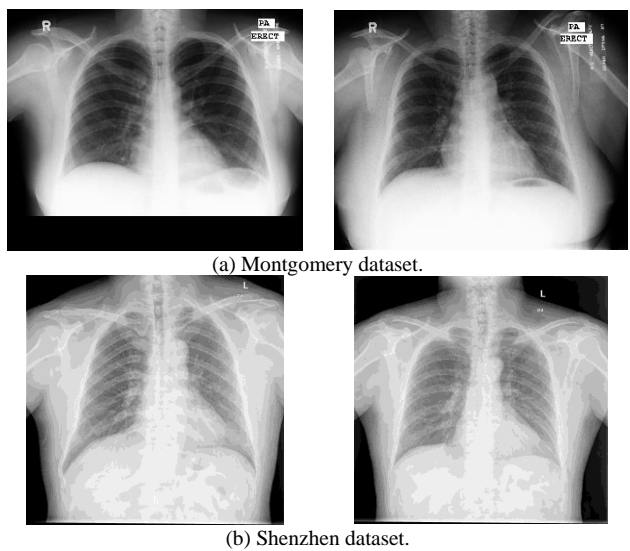

(a) Montgomery dataset.

(b) Shenzhen dataset.

Fig.1. Data visualization.

### C. Proposed Pix2pix GAN Approach

The study proposes a complete pipeline for segmenting pulmonary abnormalities on chest images of X-ray. Images and masks are loaded from the Montgomery dataset and resized, with contrast limited adaptive histogram equalization optionally employed to increase the image. This step has been taken to normalize and improve the quality of the input dataset. Fig. 3 displays the histogram variance for the lung pictures and the corresponding masks. After preprocessing, the dataset was divided into 80% training and 20% testing sets. Utilizing TensorFlow's ImageDataGenerator, applies data augmentation techniques (random rotation, shifts, flips, and zooms) to the training set. These methods increase generalization of the model and robustness.

Figure 2 illustrates the suggested architecture for the model, which is U-Net enhanced with a generative adversarial network for robust pulmonary chest X-ray abnormality detection. The GAN framework is composed of a generator and discriminator network. The generator synthesizes high-resolution images from noise vectors; it is adversarial trained to output real hallucinations of X-ray representations. Fig. 4. shows Architectural overview of the generator. At the same time, the discriminator network distinguishes between authentic and generated images, enabling the generator to create more realistic outputs, as explained in Fig. 5. The segmentation and picture quality are optimized using adversarial training. The encoder-decoder architecture of the U-Net, with skip links, is best suited for maintaining spatial information when downsampling and upsampling. The encoder gradually reduces spatial dimensions and extracts hierarchical features using cascaded 3x3 convolutional layers, ReLU activations, and max-pooling operations.

On the other hand, the decoder employs convolutional layers to reconstitute spatial resolution and predict accurately at a pixel level.

Adding a GAN adds to these capabilities by including adversarial training, which would improve the image quality and the segmentation of the images. This kind of approach is expected to work on detecting very small changes that may seem abnormal in chest X-ray scans that may be beneficial in clinical diagnostic assistance.

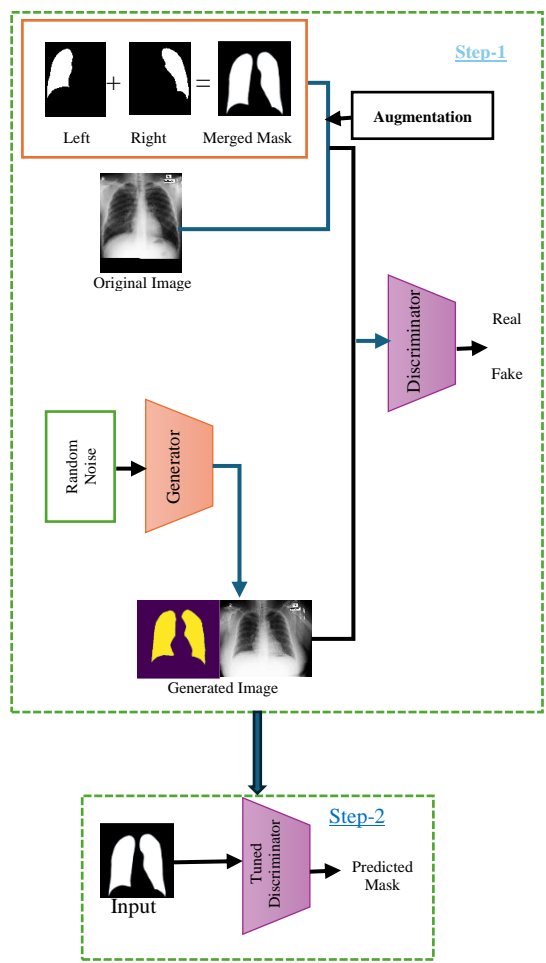

Fig.2. Proposed Pix2pix GAN Architecture.

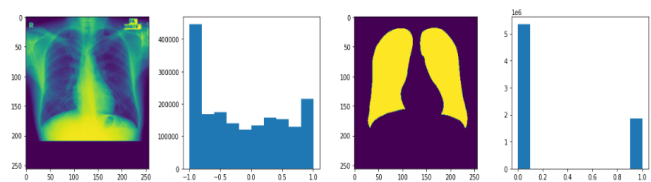

Figure 3. Histrogram difference of X-ray image and its Mask.

```
Layer (type)                 Output Shape              Param #   Connected to
=====================================================================================
input_1 (InputLayer)         [(None, 256, 256, 1)]     0
sequential_2 (Sequential)    (None, 128, 128, 64)      1024      input_1[0][0]
sequential_3 (Sequential)    (None, 64, 64, 128)       131584    sequential_2[0][0]
sequential_4 (Sequential)    (None, 32, 32, 256)       525312    sequential_3[0][0]
sequential_5 (Sequential)    (None, 16, 16, 512)       2099200   sequential_4[0][0]
sequential_6 (Sequential)    (None, 8, 8, 512)         4196352   sequential_5[0][0]
sequential_7 (Sequential)    (None, 4, 4, 512)         4196352   sequential_6[0][0]
sequential_8 (Sequential)    (None, 2, 2, 512)         4196352   sequential_7[0][0]
sequential_9 (Sequential)    (None, 1, 1, 512)         4196352   sequential_8[0][0]
sequential_10 (Sequential)   (None, 2, 2, 512)         4196352   sequential_9[0][0]
concatenate (Concatenate)    (None, 2, 2, 1024)        0         sequential_10[0][0]
                                                                 sequential_8[0][0]
sequential_11 (Sequential)   (None, 4, 4, 512)         8390656   concatenate[0][0]
concatenate_1 (Concatenate)  (None, 4, 4, 1024)        0         sequential_11[0][0]
                                                                 sequential_7[0][0]
sequential_12 (Sequential)   (None, 8, 8, 512)         8390656   concatenate_1[0][0]
concatenate_2 (Concatenate)  (None, 8, 8, 1024)        0         sequential_12[0][0]
                                                                 sequential_6[0][0]
sequential_13 (Sequential)   (None, 16, 16, 512)       8390656   concatenate_2[0][0]
concatenate_3 (Concatenate)  (None, 16, 16, 1024)      0         sequential_13[0][0]
                                                                 sequential_5[0][0]
sequential_14 (Sequential)   (None, 32, 32, 256)       4195328   concatenate_3[0][0]
concatenate_4 (Concatenate)  (None, 32, 32, 512)       0         sequential_14[0][0]
                                                                 sequential_4[0][0]
sequential_15 (Sequential)   (None, 64, 64, 128)       1049088   concatenate_4[0][0]
concatenate_5 (Concatenate)  (None, 64, 64, 256)       0         sequential_15[0][0]
                                                                 sequential_3[0][0]
sequential_16 (Sequential)   (None, 128, 128, 64)      262400    concatenate_5[0][0]
concatenate_6 (Concatenate)  (None, 128, 128, 128)     0         sequential_16[0][0]
                                                                 sequential_2[0][0]
conv2d_transpose_8 (Conv2DTrans (None, 256, 256, 1)    2049      concatenate_6[0][0]
=====================================================================================
```

Fig.4. Architectural summary of Generator.

```
Layer (type)                 Output Shape              Param #   Connected to
=====================================================================================
input_image (InputLayer)     [(None, 256, 256, 1)]     0
target_image (InputLayer)    [(None, 256, 256, 1)]     0
concatenate_7 (Concatenate)  (None, 256, 256, 2)       0         input_image[0][0]
                                                                 target_image[0][0]
sequential_17 (Sequential)   (None, 128, 128, 64)      2048      concatenate_7[0][0]
sequential_18 (Sequential)   (None, 64, 64, 128)       131584    sequential_17[0][0]
sequential_19 (Sequential)   (None, 32, 32, 256)       525312    sequential_18[0][0]
zero_padding2d (ZeroPadding2D) (None, 34, 34, 256)     0         sequential_19[0][0]
conv2d_12 (Conv2D)           (None, 31, 31, 512)       2097152   zero_padding2d[0][0]
batch_normalization_18 (BatchNo (None, 31, 31, 512)    2048      conv2d_12[0][0]
leaky_re_lu_12 (LeakyReLU)   (None, 31, 31, 512)       0         batch_normalization_18[0][0]
zero_padding2d_1 (ZeroPadding2D) (None, 33, 33, 512)   0         leaky_re_lu_12[0][0]
conv2d_13 (Conv2D)           (None, 30, 30, 1)         8193      zero_padding2d_1[0][0]
=====================================================================================
```

Fig.5. Architectural summary of Discriminator.

### D. Evaluation of the models

The proposed Pix2pix GAN model was evaluated using well-known matrices, such as the precision, Dice coefficient, recall, f1 score, and accuracy.

$$Accuracy = \frac{TP + TN}{TP + TN + FP + FN} * 100 \quad (1)$$

$$Precision = \frac{TP}{TP + FP} * 100 \quad (2)$$

$$Recall = \frac{TP}{TP + FN} * 100 \quad (3)$$

$$F1\ Score = 2 * \left(\frac{Precision * Recall}{Precision + Recall}\right) * 100 \quad (4)$$

$$Dice\ Coefficient = \frac{2 * TP}{FP + 2TP + FN} * 100 \quad (5)$$

TP = True positive, TN = True negative, FP = False positive, and FN = False negative.

## IV. OUTCOMES ANALYSIS AND DISSCUSION

The project implementation utilizes TensorFlow along with various image processing libraries such as scikit-image and OpenCV. The generator of the GAN was initially tested with test data, producing a random noise image shown in Fig. 6. Subsequently, the Montgomery dataset's training set was used to train the model. The loss of training progression is illustrated in Fig. 7, showing a consistent decrease with increasing steps. The accuracy curve, depicted in Fig. 8, shows an upward trend for both training and validation sets, converging around 99%. This indicates optimal model fitting without overfitting.

Following training, the model was evaluated using metrics presented in Table I: Accuracy (%), Precision (%), Recall (%), F1-score (%), and Dice coefficient (%), achieving high values across all metrics (98.25%, 97.92%, 98.10%, 98.01%, and 98.05% respectively). The final predicted values on the Montgomery dataset are shown in Fig. 9, demonstrating close similarity between predicted and ground truth images.

To assess robustness, the Shenzhen dataset was used to test the model., which is entirely different from the trained dataset. The accuracy curve for the Shenzhen dataset is displayed in Fig. 11, showing efficient segmentation with an accuracy of 95.87%. Predicted images are shown in Fig. 11, confirming the robustness of our proposed GAN model. To further analyze the model's response to the Shenzhen dataset, we visualized the difference between input and predicted images. Fig. 12 illustrates minimal differences between the ground truth and predicted images, affirming high accuracy in segmenting Shenzhen dataset images.

Finally, a comparison with a previous study has been summarized in Table II. The comparative table highlights that our proposed model is the first to utilize the Pix2pix GAN model for lung segmentation. We assessed its robustness using both the Montgomery and Shenzhen datasets separately. Our model achieved higher performance metrics compared to previous studies, demonstrating improvements in diversity, generality, and robustness due to the application of augmentation techniques and the superior performance capabilities of the GAN.

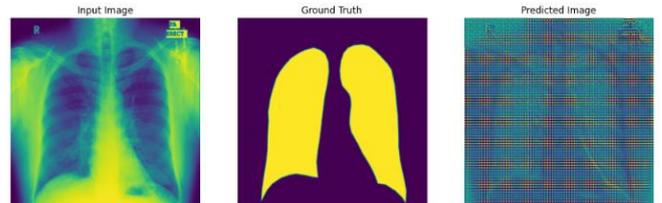

Fig.6. Predicted Image before training.

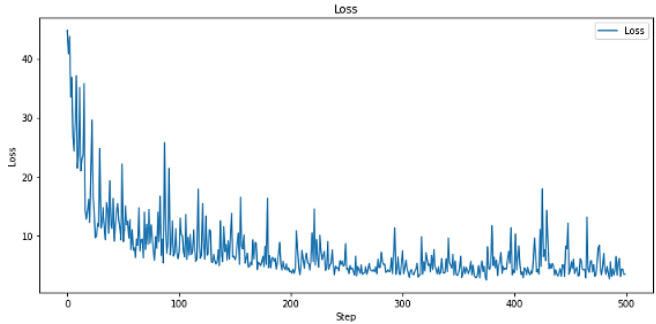

Fig.7. Training Losses.

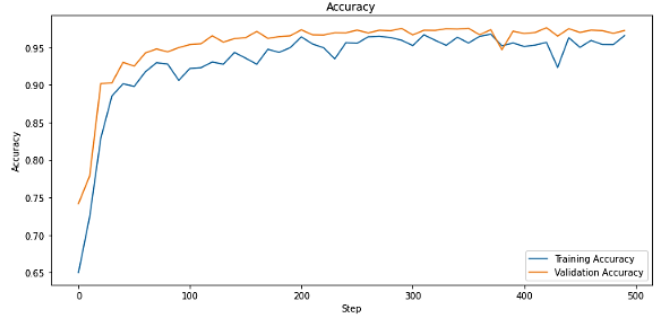

Fig.8. Accuracy Curve.

TABLE I. PERFORMANCE METRICS ON MONTGOMERY DATASET.

| Accuracy (%) | Precision (%) | Recall (%) | F1-score (%) | Dice coefficient (%) |
|---|---|---|---|---|
| 98.25 | 97.92 | 98.10 | 98.01 | 98.05 |

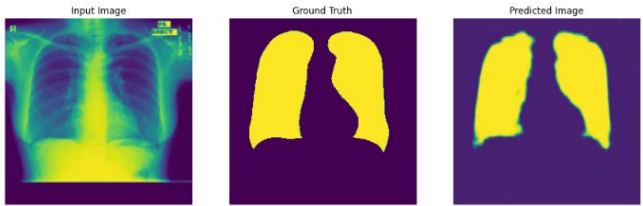

Fig.9. Predicted output of Pix2pix GAN model on Montgomery dataset.

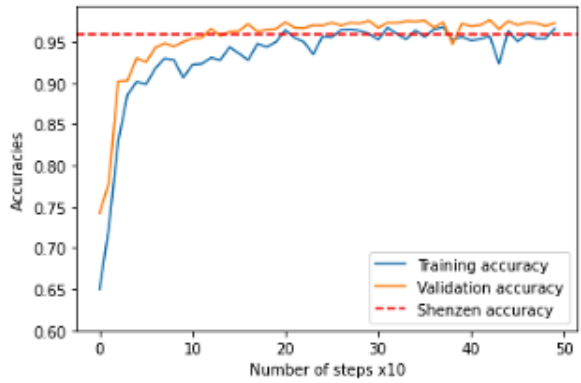

Fig.10. Accuracy curve of Shenzhen Dataset.

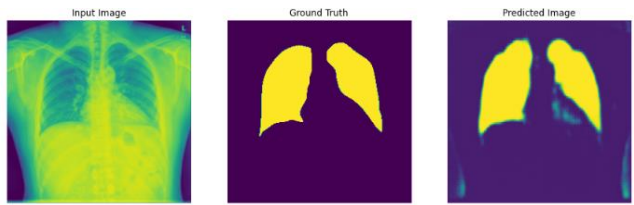

(a)

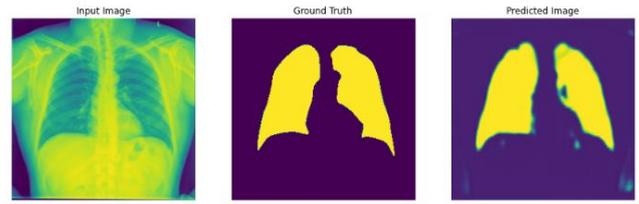

(b)

Fig.11. Predicted output of Pix2pix GAN model on Shenzhen dataset; (a) 1st part; (b) 2nd part.

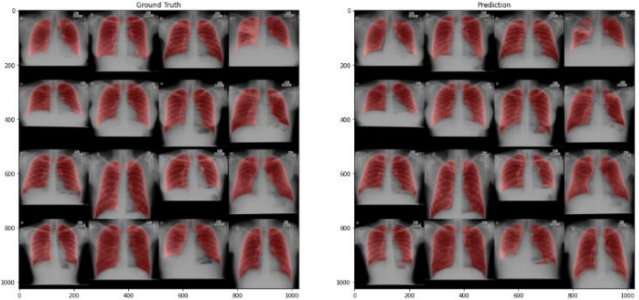

(a) Ground Truth.      (b) Predicted.

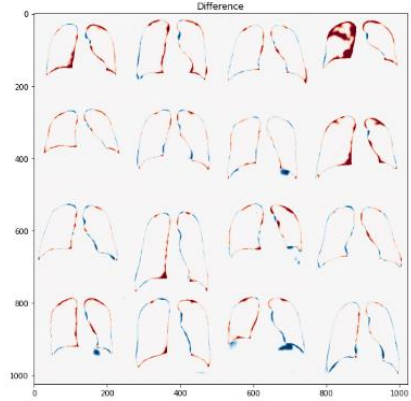

(c) Differences.

Fig.12. Visualization of the difference between ground truth and predicted output; (a) Ground truth; (b) Predicted; (c) Differences.

TABLE II. COMPARISON WITH PREVIOUS STUDY.

| Ref. | Published Year | Model | Robustness Check | Performance |
|---|---|---|---|---|
| [10] | 2024 | Pulmon U-Net | No | Accurcay= 94.25%. |
| [14] | 2024 | LSEG | No | Accuracy= 97.12% |
| [16] | 2023 | SA-UNet | NO | D.cof. = 96.80% IoU= 91.97% |
| [11] | 2023 | Unet | No | Accuracy = 97.34% |
| Our Proposed Approach | - | Pix2pix GAN | Yes Trained by Montgomery and Tested by Shenzhen | Accuracy=98.25 %, F1-score =98.01% Robusness Accuracy=95.87 % |

## V. CONCLUSION AND FUTURE WORK

Finally, our work offers a new method for segmenting pulmonary chest X-ray abnormalities employing a Pix2pix GAN model. We have rigorously experimented with it on the Montgomery and Shenzhen datasets, demonstrating its usefulness with better performance metrics than earlier approaches. The special abilities of GANs and data augmentation approaches have enhanced our model's strength, which exhibits promising advances in clinical diagnostic aid. By using advanced machine-learning approaches, this research improves the accuracy of segmentation and sets a standard for future advancements in medical image analysis.

A possibility for improvement in this line of research is the limited number of datasets considered. Thus, we focused more on the Montgomery and Shenzen datasets and did not explore reactions to an extended list of datasets here. Offspring analysis could create this strategy and confirm and increase the viability of our advised procedure by developing a greater extent of various information. This expansion may increase the model's applicability to clinical settings and situations.